\documentclass[%
aip,%
jmp,%
reprint,%
onecolumn,
A4paper,10pt]{revtex4-1}

\usepackage{amsfonts}							
\usepackage{amssymb}							
\usepackage{amsmath}							
\usepackage{bbm}								
\usepackage{bm}									
\usepackage{mathrsfs}							
\usepackage{eucal}								
\usepackage{color}								
\usepackage[dvipsnames]{xcolor}								
\usepackage[dvips,pdftex]{graphicx}		
\usepackage{setspace}
\usepackage[ pdftex, plainpages = false, pdfpagelabels,
                 pdfpagelayout = SinglePage,
                 bookmarks,
                 bookmarksopen = true,
                 bookmarksnumbered = true,
                 breaklinks = true,
                 linktocpage,
                 colorlinks = true,
                linkcolor = NavyBlue,
                 urlcolor  = RoyalPurple,
                 citecolor = ForestGreen,
                 anchorcolor = Green,
                 hyperindex = true,
                 hyperfigures
                 ]{hyperref} 					

\usepackage[left=1.2in, right=1.2in, top=1in, bottom=1in, headheight=16.9pt]{geometry}	

\newcommand{\bt}			{\beta}

\newcommand{\dlt}			{\delta}

\newcommand{\tta}			{\theta}

\newcommand{\kp}			{\kappa}
\newcommand{\lm}			{\lambda}

\newcommand{\vfi}			{\varphi}
\newcommand{\ro}			{\rho}
\newcommand{\vro}			{\varrho}

\newcommand{\Gm}			{\Gamma}
\newcommand{\Dlt}			{\Delta}
\newcommand{\Tta}			{\Theta}

\newcommand{\Ups}		{\Upsilon}
\newcommand{\Om}			{\Omega}

\newcommand{\mc}[1]{\mathcal{ #1}}						
\newcommand{\trm}[1]{\textrm{ #1}}						
\newcommand{\mf}[1]{\mathfrak{ #1}}						
\newcommand{\mb}[1]{\mathbf #1}			  				
\newcommand{\ph}[1]{\phantom{ #1}}			  				

\newcommand{\wt}[1]{\widetilde{ #1}}						
\newcommand{\wht}[1]{\widehat{ #1}}						

\newcommand{\ud}{\mathrm{d}} 								

\newcommand{\R}{\mathbbm{R}}								
\newcommand{\C}{\mathbbm{C}}								

\newcommand{\Ad}[1]{\mathrm{Ad}_{#1}\,}				
\newcommand{\I}{\mathbbm{1}}								

\newcommand{\x}{\wedge}																

\newcommand{\half}{\frac{1}{2}}								

\newtheorem{theorem}{Theorem}[section]					
																				
\newcommand{\qed}{\nobreak \ifvmode \relax \else			
      \ifdim\lastskip<1.5em \hskip-\lastskip							
      \hskip1.5em plus0em minus0.5em \fi \nobreak
      \vrule height0.75em width0.5em depth0.25em\fi}
      
\newcommand{\trian}{\nobreak \ifvmode \relax \else			
      \ifdim\lastskip<1.5em \hskip-\lastskip
      \hskip1.5em plus0em minus0.5em \fi \nobreak
      \vrule height0.75em width0.5em depth0.25em\fi}
			
\newcommand{\tbf}[1]{\textbf{#1}}														

\newcommand{\reff}[1]{~(\ref{#1})}															
\newcommand{\citte}[1]{~\cite{#1}}															

\newcommand{\cf}{\emph{cf.\,}}															

\newcommand{\hfis}{\mathcal{H}_{\!\!{\trm{\,phys}}}}						
\newcommand{\haux}{\mathcal{H}_{\!\!{\trm{aux}}}}							
\newcommand{\thaux}{{\widetilde{\mathcal{H}}_{\!\!{\trm{aux}}}}}		
							 
\newcommand{\aux}[2]{\left( #1, #2\right)_{\!\!\trm{aux}}}				
\newcommand{\taux}[2]{\left( #1, #2\right)_{\wt{\text{aux}}}}			
\newcommand{\brs}[2]{\left( #1, #2\right)_{\text{BRST}}}					
					
\newcommand{\BbbR}{\mathbb{R}}

\newcommand{\f}[3]{{f_{#1#2}}^{#3}}												
\newcommand{\ga}[2]{\left( #1, #2\right)_{\!\trm{{ga}}}}				
\newcommand{\tga}[2]{\left( #1, #2\right)_{\wt{\trm{\!{ga}}}}}		
\newcommand{\reg}[2]{\left( #1, #2\right)_{\!\trm{\small\textsc{brst}}}^{\vro}}
\newcommand{\dlpartial}[2]{\dfrac{\partial^{\,\ell} #1}{\partial #2}}	

\numberwithin{equation}{section}				

\begin{document}


\title{Refined algebraic quantisation in a system with nonconstant gauge invariant structure functions} 



\author{Eric Mart\'inez-Pascual}
\thanks{emartinez@fcfm.buap.mx}
\affiliation{\footnotesize{ School of Mathematical Sciences, University of Nottingham, Nottingham NG7 2RD, United Kingdom}}
\affiliation{\footnotesize{ Facultad de Ciencias F\'{i}sico Matem\'{a}ticas, Benem\'{e}rita Universidad Aut\'{o}noma de Puebla,}
\footnotesize{Apartado Postal 1152, Puebla, Puebla, M\'{e}xico.}}


\date{29 October 2013}

\begin{abstract}
In a previous work [J. Louko and E. Mart\'{i}nez-Pascual, ``Constraint rescaling in refined algebraic quantisation: Momentum constraint,'' J. Math. Phys. \tbf{52} 123504 (2011)], refined algebraic quantisation (RAQ) within a family of classically equivalent constrained Hamiltonian systems that are related to each other by rescaling \emph{one} momentum-type constraint was investigated. In the present work, the first steps to generalise this analysis to cases where more constraints occur are developed. The system under consideration contains \emph{two} momentum-type constraints, originally abelian, where rescalings of these constraints by a non-vanishing function of the coordinates are allowed. These rescalings induce structure functions at the level of the gauge algebra. Providing a specific parametrised family of real-valued scaling functions, the implementation of the corresponding rescaled quantum momentum-type constraints is performed using RAQ when the gauge algebra: $(i)$ remains abelian and $(ii)$ undergoes into an algebra of a nonunimodular group with nonconstant gauge invariant structure functions. Case $(ii)$ becomes the first example known to the author where an open algebra is handled in refined algebraic quantisation. Challenging issues that arise in the presence of non-gauge invariant structure functions are also addressed.
\end{abstract}


\maketitle 

\section{Introduction}
\label{sec:intro} 

In the mentioned previous work\citte{lou11} (which will be abbreviated by Paper I below), we studied refined algebraic quantisation (RAQ) within a family of classically equivalent constrained Hamiltonian systems that are related to each other by rescaling \emph{one} momentum-type constraint. We found that depending on the asymptotics of the scaling function, the quantum implementation of the constraint: (i) is equivalent to that with identity scaling; (ii) fails, due to the non-existence of self-adjoint extensions of the constraint operator; and (iii) shows an ambiguity which arises from the self-adjoint extension of the constraint operator, and the resolution of this purely quantum mechanical ambiguity determines a superselection structure of the physical Hilbert space. In the present paper this analysis is partly extended by including another momentum-type constraint into the model.

In a bosonic gauge system with one momentum-type constraint,  due to the antisymmetry of the Poisson bracket (PB), the rescaling of the constraint by an invertible function on the phase space  does not change the abelian nature of the gauge algebra. The story is, however, quite different in a model with at least two constraints which close under the PB with structure constants (\emph{closed} algebra); redefining the constraints by an invertible linear map that is not constant on the phase space yields PBs between the constraints that can be arranged to close with nonconstant structure functions (\emph{open} algebras), see, for example\citte{kuc86a}. At a classical level, the presence of these artificially constructed structure functions changes the classification of the gauge algebra, from a closed to an open algebra, but the Dirac observables remain invariant.

At a quantum level, a broad proposal for implementing phase space constraints was given by Dirac\citte{dirbk-lec}. RAQ is a precise formulation of Dirac's programme and its subtleties. In RAQ one establishes an auxiliary Hilbert space $ \haux $ on which the constraints will act as self-adjoint operators; observables and the physical state condition are interpreted in terms of distributions on some dense subspace $ \Phi\subset\haux $. Within this scheme, the concept of a rigging map ($ \eta $) is axiomatically introduced to solve the constraints and, simultaneously,  define an inner product between  physical states. The physical Hilbert space  $ \hfis $ is constructed by completion\citte{hig91a, hig91b, ash95, mar00}. The group averaging procedure has shown to be an effective way to construct a rigging map\citte{giu99a, giu99b, Giulini:1999kc}, with a vast amount of successful applications found in the literature\citte{lou11, gom99, Louko:1999tj, Marolf:2008it, Marolf:2008hg, Louko:2003cn, Louko:2004zq, lou05}. Closed gauge algebras are desirable in order to apply this technique: a Lie group action generated by the quantum constraints is expected in such cases. Since closed gauge algebras can always be turned into open ones, it results interesting to investigate the averaging techniques for open algebras with a genuine Lie group action underlying them. Rescaling two momentum-type constraints can be considered as a good laboratory for this.

The implementation of constraints at a quantum level is highly affected by the mathematical structure of the quantum gauge algebra. An explicit example is found in the variation of Dirac's condition for physical states implied by quantum constraints that generate a non-unimodular Lie group\citte{giu99a, duv89, tuy90, duv91}. Using Becchi-Rouet-Stora-Tyutin (BRST) formalism, in its Hamiltonian form\citte{bat77, henrep}, a proposal for extending the group averaging technique to open gauge algebras has been given by Shvedov\citte{shv02} based on previous works by Batalin and Marnelius\citte{bat95}. In Paper I, delicate subtleties of this proposal  have been remarked such as the inclusion of a test space, technical details on the self-adjointness of the quantum rescaled constraint operator, and the sense in which the averaging acts and converges. The presence of a single constraint, however, limited us from testing the proposal when actual structure functions are present. In this paper this gap is filled by analysing two momentum-type constraints rescaled with nonvanishing functions on the configuration space. 

The paper is organised as follows. The classical system is introduced in Sec.~\ref{the-model}. In order to have control over the infinite number of possible algebras obtained by an arbitrary rescaling of the two constraints, a specific parametrised family of real-valued scaling functions will be used and it is described in Sec.~\ref{sec:scaling family}. Depending on the values taken by the parameters, either the gauge algebra $(i)$ is maintained, $(ii)$ becomes an open algebra with gauge invariant structure functions that generates a non-unimodular Lie group action at each point in the physical configuration space, or $(iii)$ becomes a full open algebra --that is, the artificial structure functions depend on all configuration variables. Sec.~\ref{sec:raq} is devoted to the analysis of RAQ to cases $(i)$ and $(ii)$, it turns out that the family of scaling functions in these two cases belongs to the Type I as categorised in Paper I. In Sec.~\ref{sec:final comm}, final remarks and general comments on the more involved case $(iii)$ are presented. As the new ingredient of nonconstant gauge invariant structure functions emerges in case $(ii)$, based on Shvedov's proposal, in Appendix~\ref{app:BRSTanalysis} a `group averaging' ansatz for these type of open gauge algebras is derived. In Appendix~\ref{app:gauge group}, basic properties of the gauge group generated by the open algebra with gauge invariant structure functions of case $(ii)$ are placed.

\section{Classical system: two momentum-type constraints}
\label{the-model}

An immediate generalisation of the system considered in Paper I is that with configuration space 
$\BbbR^3 = \{(\tta,\vfi,x)\}$ and phase space 
$\Gm = T^*\BbbR^3 = \{(\tta,\vfi, x, p_{\tta},p_{\vfi}, p_x)\} \simeq \BbbR^6$ constrained by 
\begin{subequations}\label{eq:rcs}
\begin{align}
\phi_{1} & := M(\tta,\vfi,x)\,p_{\tta} \approx 0\ ,\label{eq:rcs1}\\
\phi_{2} & := N(\tta,\vfi,x)\,p_{\vfi} \approx 0\ .\label{eq:rcs2}
\end{align}
\end{subequations}

The scaling functions $ M $ and $ N $ are assumed to be real-valued,  nonvanishing, smooth and positive functions on the configuration space.  These conditions ensure that the set of constraints  \eqref{eq:rcs1} and  \eqref{eq:rcs2} is regular  and irreducible in the sense described in Ref.~\cite{henbk}. The constraint surface $ \Gm_{c} $ defined by\reff{eq:rcs} coincides with the constraint surface defined by the corresponding \emph{un}scaled constraints $ p_{\tta} \approx 0 $ and $ p_{\vfi} \approx 0 $.  The vector fields $ X_{1}:=M(\tta, \vfi, x)\,\partial_{\tta} $ and $ X_{2}:= N(\tta,\vfi, x)\, \partial_{\vfi} $ are linearly independent at each point of the constraint surface; moreover, depending on the nature of the scaling functions, the vectors fields $ X_{a} $ may or may not be complete on the phase space. It will be assumed that there is no true Hamiltonian, that is, the restriction of the so-called total Hamiltonian\cite{henbk} to $ \Gm_{c} $ vanishes; a true Hamiltonian, if any, can only depend on the true degree of freedom $x$ and its canonical pair $p_x$, inclusion of a true Hamiltonian would be straightforward.

In contrast to the abelian gauge algebra formed by the unscaled constraints $ p_{\tta} \approx 0 $ and $ p_{\vfi} \approx 0 $, for the set of constraints\reff{eq:rcs} structure functions appear at the gauge algebra level, explicitly
\begin{equation}\label{eq:gauge algebra}
\{\phi_{a},\phi_{b}\}=\f{a}{b}{c}(q)\phi_{c}\ ,\quad (a=1,2)\ ,
\end{equation}
with
\begin{subequations}\label{eq:structure fncs}
\begin{align}
\f{1}{1}{a} & =0=\f{2}{2}{a} \ ,  \\
\f{1}{2}{1} & = N(\partial_{\vfi}\ln M)\ , \\
\f{1}{2}{2} & = - M(\partial_{\tta}\ln N)\ .
\end{align}
\end{subequations}
This process of rescaling constraints is a prototype of the way one can turn any genuine Lie gauge algebra, in this case an abelian one, into an open gauge algebra\reff{eq:gauge algebra}. Rescalings of these type are not harmful at the classical level. Dirac observables of the theory are still functions on the reduced phase space $ \Gm_{\!\!\trm{{red}}}=\{(x,p_{x})\}\simeq\R^{2} $. In subsection \ref{sec:scaling family}, a specific family of scaling functions, which covers the spectrum of possibilities present when two momentum-type constraints are rescaled with functions on the configuration space, is given.

Rescaling constraints is not the only way in which one can produce open gauge algebras from closed ones; for instance, from a set of reducible constraints, whose PBs close with structure constants, one can extract a linearly independent subset of gauge generators whose PBs close with nonconstant structure functions\citte{kuc86a}.

\subsection{Rescaling two momentum constraints: A family of scaling functions}\label{sec:scaling family}

Consider now a particular choice of the scaling functions $ M $ and $ N $, such that each of them does not depend on the associated coordinate to the momentum it is rescaling, that is, $ M(\vfi,x) $ and $ N(\tta,x) $, respectively. To be  precise, let $ M(\vfi,x)$  be $ f(x) e^{\kp_{1}\vfi} $, and $ N(\tta,x) $ be $ g(x)e^{\kp_{2}\tta} $, hence,\reff{eq:rcs1} and  \eqref{eq:rcs2} descend into
\begin{subequations}\label{eq:M and N}
\begin{align}
\phi_{1} & =M(\vfi,x) p_{\tta}=f(x) e^{\kp_{1}\vfi}p_{\tta}\approx 0\ ,\label{eq:M}\\
\phi_{2} & =N(\tta,x)p_{\vfi}=g(x)e^{\kp_{2}\tta}p_{\vfi} \approx 0\ , \label{eq:N}
\end{align}
\end{subequations}
with $ \kp_{a} $ real-valued parameters, and functions $ f $ and $ g $ consistent with the conditions on the generic scaling functions $ M $ and $ N $ of Eqs.~\eqref{eq:rcs}, that is, $ f $ and $ g $ are assumed to be smooth, real--valued, nonvanishing, and positive functions on $ \R $. Worth noticing is the particular dependence of $ M $ and $ N $ as $ M(\vfi,x) $ and $ N(\tta,x) $, respectively; at a classical level this implies that the constraint vector fields $ X_{\tta}:=  M(\vfi,x) \partial_{\tta}$ and $ X_{\vfi}:= N(\tta,x)\partial_{\vfi} $ are complete on the constraint surface. It is this classical property whose implication will be the existence of self-adjoint constraint operators at a quantum level.

The nonzero structure functions in the algebra correspond to
\begin{subequations}\label{eq:family structure fncs}
\begin{align}
\f{1}{2}{1} & = \kp_{1}\,g(x)e^{\kp_{2}\tta}\ , \label{eq:fncs1}\\
\f{1}{2}{2} & = -\kp_{2}\,f(x)e^{\kp_{1}\vfi}\ .  \label{eq:fncs2}
\end{align}
\end{subequations}

From appropriate choices of the real parameters, $ \kp_{a} $, three qualitatively different gauge algebras emerge:
\begin{enumerate}
\item[$(i)$]\emph{Abelian gauge algebra}:  Case $\kp_{1}=0=\kp_{2}$.  This corresponds to the case of rescaling the abelian constraints $p_{\tta}\approx0$ and $p_{\vfi}\approx 0$ with the nonvanishing gauge invariant functions $f(x)$ and $g(x)$ respectively. Such rescalings do not change the abelian nature of the unscaled constraints
\begin{equation*}
\{\phi_{1},\phi_{2}\}=0\ .
\end{equation*}

\item[$(ii)$]\emph{Gauge invariant structure functions}: Case either $\kp_{1}$ or $\kp_{2}$ vanishes. Here the structure functions\reff{eq:family structure fncs} only depend on the physical degree of freedom $x$. For instance, choosing $\kp_{2}=0$ and $\kp_{1}\equiv\kp\neq 0$, the only nonzero ($x$-de\-pen\-dent) structure function is 
\begin{equation}\label{eq:xdependent structure func}
\f{1}{2}{1}  = \kp g(x)\ .  \\
\end{equation}
The open gauge algebra is $\{\phi_{1},\phi_{2}\}=\kp g(x)\phi_{1}$. The interpretation of the gauge group in this case is immediate: At a fixed point $x$ the gauge group is a nonunimodular triangular subgroup of $ GL(2,\R) $ (see Appendix~\ref{app:gauge group}). If the function $ g $ is constant everywhere, a genuine gauge Lie algebra structure is recovered.

\item[$(iii)$]\emph{Non gauge invariant structure functions}:  Case $\kp_{1}\neq 0$, $\kp_{2}\neq 0$. One obtains a set of first-class constraints where the structure functions depend on all the configuration variables.

\end{enumerate}
The family of scaling functions introduced in\reff{eq:M and N} hence embraces the cases where the gauge algebra of the unscaled constraints $p_{\tta}\approx0$ and $p_{\vfi}\approx 0$ either is kept abelian as in case $ (i) $ or contains artificial structure functions as in cases $ (ii) $ and $ (iii) $.  In spite of the restrictive family of scaling functions under consideration, \cf~Eqs.~\eqref{eq:M and N}, it turns out that the associated constraint algebra may be as simple as in case $ (i) $, yet embrace the general situation of having nonvanishing structure functions which depend on all coordinates as in case $ (iii) $.

 If the unscaled constraints had had a gauge algebra different from the abelian one, gauge invariant scaling functions would have introduced gauge invariant structure functions at the level of the gauge algebra.  
 
\section{Refined algebraic quantisation}\label{sec:raq}

Following RAQ as reviewed in Ref.\citte{mar00}, in this section the system of constraints\reff{eq:M and N} for the particular cases $ (i) $ and $ (ii) $ is quantised. General comments on case $ (iii) $ are given in Sec.~\ref{sec:final comm}. In the subsection \ref{sec:as} the Hilbert space $ \haux $ required by RAQ is set.   In subsection \ref{sec:rm-i}, the quantisation of the case $ (i) $ is analysed. This case falls into the range of applicability of the group averaging method\citte{mar00, giu99a}, a powerful technique for constructing a rigging map which simultaneously solves the constraints and defines an inner product in the physical Hilbert space $ \hfis $. In subsection \ref{sec:rm-ii}, quantisation of case $ (ii) $ is treated, as it includes the new ingredient of structure functions in the gauge algebra, a new type of `group averaging' ansatz must be adopted. The regularised BRST inner product for gauge systems\citte{shv02, Marnelius:1990eq, mar94a, mar94, mar93a, bat95} is employed to obtain it (see also Appendix \ref{app:BRSTanalysis}).

\subsection{Auxiliary Hilbert Space}\label{sec:as}

The space of complex-valued square integrable functions on the classical configuration space , $ \haux:=L^{2}(\R^{3},\ud\tta\ud\vfi\ud x) $, is chosen as the auxiliary Hilbert space. The space $ \haux $ is equipped with the auxiliary inner product
\begin{equation}\label{eq:aux inner}
\aux{\psi}{\chi}:=\int_{\R^{3}}\!\ud^{3} q\;{\psi^{*}(q)}\chi(q)\ ,
\end{equation}
where $ \ud^{3} q $ represents the Lebesgue measure, $ q $ stands for all the coordinates $(\tta,\vfi,x) $, and $ {}^{*} $ denotes complex conjugation. The basic quantum operators $ \wht{q} $ act by multiplication and the momentum operators $ \wht{p}:=-i\partial_{q} $ act by derivation on elements in $ \haux $. These operators become self-adjoint on the dense subspace $ C^{\infty}_{0}(\R^{3}) $, the set of all smooth functions  with compact support in $ \haux $. 

\subsection{RAQ for case $ (i) $}\label{sec:rm-i}
 
 For case $ (i) $,   that is $ \kp_{a}=0 $, the classical constraints Eqs.~(\ref{eq:M and N}) are promoted as  the following operators on $ \haux $:
\begin{subequations}\label{eq:c ops i}
\begin{align}
\wht{\phi}_{1}:= & f(x)\wht{p}_{\tta}\label{eq:fi1 operator i}\ ,\\
\wht{\phi}_{2}:= & g(x)\wht{p}_{\vfi}\label{eq:fi2 operator i}\ ,
\end{align}
\end{subequations}
where no ordering problems were encountered. The quantum gauge algebra is abelian: $ [\wht{\phi}_{1},\wht{\phi}_{2}]=0 $.  The constraint operators become self-adjoint on $ C^{\infty}_{0}(\R^{3}) $. The unitary action of the gauge group is then 
\begin{equation}\label{eq:Uaction i}
(\wht{U}(\lm^{a})\psi)(\tta,\vfi,x):=\psi(\tta +f(x)\lm^{1},\vfi+g(x)\lm^{2},x)\ ,
\end{equation}
which shows that $ \wht{U} $ takes elements from $ C^{\infty}_{0}(\R^{3})  $ onto $ C^{\infty}_{0}(\R^{3}) $.  

As an abelian gauge algebra falls in the range of applicability of the group averaging formula reported in Refs.\citte{giu99a, mar00}, the following sesquilinear form is introduced:
\begin{equation}\label{eq:ga i}
\ga{\psi}{\chi}:=\int\ud^{2}\lm\;\aux{\psi}{\wht{U}(\lm^{a}) \chi}\ .
\end{equation}
From\reff{eq:Uaction i}, the multiplication law of the gauge group is $ \wht{U}(\lm^{a})\wht{U}(\lm'^{a})=\wht{U}(\lm^{a}+\lm'^{a}) $, therefore, $ \{\wht{U}(\lm^{a}):\:\lm^{a}\in\R\}\simeq \R^{2} $. Hence, the range of integration in this integral will be taken to be the whole $ \R^{2} $. It is worth noticing that in the case where $ f_{ab}^{\ph{ab}c}=0 $, for all values of $ a,b $ and $ c $, the regularised BRST  inner product\reff{eq:final reg inner} duly reduces to the group averaging formula\reff{eq:ga i}.

A more convenient way to express the group averaging formula\reff{eq:ga i} is gained by mapping the Hilbert space $ \haux $ into $ \thaux:=L^{2}(\R^{3},\ud\Tta\ud\Xi\ud x)  $ through the following isomorphism:
\begin{align}
\haux & \to \thaux \ , \notag\\
\psi & \mapsto \wt{\psi} \ , \notag\\
\wt{\psi}(\Tta,\Xi,x) & := \sqrt{f(x)g(x)} \, \psi\bigl(\Tta f(x),\Xi\,g(x),x\bigr)\ . \label{eq:haux to thaux 2c}
\end{align}
The Hilbert space  $ \thaux $ is endowed with the positive definite inner product
\begin{equation}\label{eq:thaux inner 2}
\taux{\wt{\psi}}{\wt{\chi}}:=\int_{\R^{3}}\!\ud\Tta\ud\Xi\ud x\;{\wt{\psi}^{*}(\Tta,\Xi,x)}\wt{\chi}(\Tta,\Xi,x)\ .
\end{equation}
Therefore, the group averaging sesquilinear form Eq.\reff{eq:ga i} on $ \thaux $ becomes 
\begin{equation}\label{eq:tga i}
\tga{\wt{\psi}}{\wt{\chi}}=\int_{\R^{2}}\ud^{2}\lm\;\taux{\wt{\psi}}{\wt{U}(\lm^{a}) \wt{\chi}}\ ,
\end{equation}
where the action of the gauge group on elements in $ \thaux $ explicitly reads 
\begin{equation}\label{eq:tuaction 2c}
\left(\wt{U}(\lm^{a}) \wt{\chi}\right)(\Tta,\Xi,x)=\wt{\chi}(\Tta+\lm^{1},\Xi+\lm^{2},x)\ .
\end{equation}
The system has then been mapped to that in which both functions $ f $ and $ g $ are the constant function 1. 

Therefore, RAQ can be carried out in $ \thaux $  as discussed in Sec. IIB of Ref.\citte{ash95}. One chooses the smooth functions of compact support on $ \R^{3}=\{(\Tta,\Xi, x)\} $ to be the linear dense subspace $ \Phi $ of test states required by RAQ; on this space, the integral\reff{eq:tga i} is well defined. For any $ \wt{\psi}\in \Phi $, the antilinear map $ \eta $ from $ \Phi $ to the algebraic dual $ \Phi' $ is defined by
\begin{equation}\label{eq:rigg map i}
\eta[\wt{\psi}](\wt{\chi}):=\tga{\wt{\psi}}{\wt{\chi}} .
\end{equation}
A Fourier transformation shows that the averaging projects out all gauge dependence from the wave functions and that the physical Hilbert space becomes $ L^{2}(\R,\ud x) $. In the classification of scaling functions given in Paper I, this case falls into the type I scaling functions category. 

\subsection{RAQ for case $ (ii) $}\label{sec:rm-ii}
In this subsection, the following system of first class constraints will be examined:
\begin{subequations}\label{eq:constraints ii}
\begin{align}
\phi_{1}:= & f(x)e^{\kp\vfi}p_{\tta}\approx 0\label{eq:fi1 ii}\ ,\\
{\phi}_{2}:= & g(x)p_{\vfi}\approx 0\label{eq:fi2 ii}\ .
\end{align}
\end{subequations}
The gauge algebra in this case presents nonconstant gauge invariant structure functions, 
\begin{equation}\label{eq:gauge algebra ii}
\{\phi_{1},\phi_{2}\}=f_{12}^{\ph{12}1}(x)\phi_{1}=\kp g(x)\phi_{1}\ .
\end{equation}

In the quantum theory, the construction of the constraint operators $ \wht{\phi}_{a} $ does not involve any ordering issues and they read as
\begin{subequations}\label{eq:c ops ii}
\begin{align}
\wht{\phi}_{1}:= & \ f(x)e^{\kp\vfi}\wht{p}_{\tta}\label{eq:fi1 op ii}\ ,\\
\wht{\phi}_{2}:= & \ g(x)\wht{p}_{\vfi}\label{eq:fi2 op ii}\ .
\end{align}
\end{subequations}
They obey the non-anomalous quantum algebra of constraints
\begin{equation}\label{eq:comm of constraints ii}
\big[\,\wht{\phi}_{1},\wht{\phi}_{2}\,\big]=i\kp g(x)\wht{\phi}_{1}\ .
\end{equation}

These constraint operators are symmetric with respect to the auxiliary inner product\reff{eq:aux inner} on the dense subspace $ C^{\infty}_{0}(\R^{3}) $. Using the basic criterion for self-adjointness by Von Neumann,  it is not difficult to see that each pair of deficiency indices  $ (n_{+},n_{-}) $ associated to each constraint operator $ \wht{\phi}_{a} $ is $ (0,0) $; the constraint operators are hence self-adjoint. The norm of the solutions to $ \wht{\phi}_{a}\psi =\pm i\psi $ for each $ a $ diverges  because of the behaviour either at $ \tta\to\infty $ or at $ \tta\to-\infty $.

The algebra of the quantum constraints\reff{eq:comm of constraints ii} is in one-to-one correspondence with the generators of the group $B(2,\R) $ at each point $ x $.  In the  Appendix~\ref{app:gauge group}, basic properties of this gauge group are placed. Hence, exponentiation of Eqs.\reff{eq:c ops ii} yields a unitary representation  $ \wht{U} $, at each point $ x $, of $B(2,\R) $ on $ \haux $. The group elements that appear in the decomposition\reff{eq:exp3} are represented on $ \haux $ as
\begin{subequations}\label{eq:reps of bl2r}
\begin{align}
\wht{U}\Big[\exp\big(\bt\,T_{1}(x)\big)\Big] = &  \ \exp\left(i\bt\wht{\phi}_{1}\right)  \label{eq:eT1}\\
\wht{U}\left[\exp\left(\dfrac{\lm^{2}}{2}\,T_{2}(x)\right)\right] = & \ \exp\left(i\dfrac{\lm^{2}}{2}\wht{\phi}_{2}\right)  \label{eq:eT2}
\end{align}
\end{subequations}
where $ \bt$ is given by
\begin{equation*}
\bt:=\lm^{1}\, \dfrac{\sinh\left(\half\kp g(x)\lm^{2}\right)}{\half\kp g(x)\lm^{2}}\ .
\end{equation*}

Each operator in Eqs.\reff{eq:reps of bl2r} acts on functions in $ \haux $ as follows:
\begin{subequations}\label{eq:bl2r action}
\begin{align}
\left[\exp\left(i\bt\wht{\phi}_{1}\right)\psi\right](\tta,\vfi,x)  = & \ \psi(\tta+\bt f(x) e^{\kp\vfi},\vfi,x)\ ,\label{eq:eT1 action}\\
\left[\exp\left(i\dfrac{\lm^{2}}{2}\wht{\phi}_{2}\right)\psi\right](\tta,\vfi,x)  = & \ \psi(\tta,\vfi+\lm^{2} g(x)/2,x)\ . \label{eq:eT2 action}
\end{align}
\end{subequations}
Hence, the action of a general element of $ B(2,\R) $ on a wave function produces a shift in the $ \vfi $-direction by $ \half\lm^{2}\,g(x) $, see\reff{eq:eT2 action}, which is followed by a shift in the $ \tta $-direction by $ \bt f(x)e^{\kp\vfi} $, see\reff{eq:eT1 action}, and ends with another shift  in the $ \vfi $-direction by $ \half\lm^{2}\,g(x) $. The final result being
\begin{equation}\label{eq:Uaction 2b}
\left(\wht{U}(g)\psi\right)(\tta,\vfi,x)=\psi\left(\tta+ \bt f(x)e^{\kp\vfi},\,\vfi+\lm^{2} g(x),\,x\right) \ .
\end{equation}

We wish to use the `group averaging' ansatz\reff{eq:final reg inner}, which allows $ x $-dependent structure functions, and obtain a physical inner product. A direct calculation shows that the $ x $-dependent symmetric measure in\reff{eq:final reg inner} (see also\reff{eq:symm measure}) is in this case
\begin{equation}\label{j0 case 2b}
\vert j_{0}(\mb{u}(x,\lm)) \vert\ud^{2}\lm=\dfrac{\sinh\left[\half \lm^{2} \kp g(x)\right]}{\half\lm^{2}\kp g(x)}\ud^{2}\lm\ .
\end{equation}
This positive quantity  coincides with the symmetric measure $ \ud_{0}g(x) $ independently obtained in Appendix~\ref{app:gauge group}, see\reff{eq:symm measure b2r}. Therefore, the group averaging ansatz reads
\begin{equation}\label{eq:ga case 2b}
\int\ud^{2}\lm\,\ud^{3}q\:\dfrac{\sinh\big[\half\lm^{2}\kp g(x)\big]}{\half\lm^{2}\kp g(x)}\,{\psi^{*}(q)}\big(\wht{U}(g)\chi\big)(q)\ .
\end{equation}

In order to make sense of this formula,  it is convenient to map $ \haux $  to $ \thaux $ via the following isomorphism:
\begin{align}
\haux & \to \thaux\ , \notag\\
\psi & \mapsto \wt{\psi}\ , \notag\\
\wt{\psi}(\Tta,\Xi,x) & := \sqrt{f(x)g(x)e^{\kp \vfi}} \, \psi\bigl(\Tta f(x)e^{\kp\vfi},\Xi\,g(x),x\bigr)\ . \label{eq:haux to thaux 2b}
\end{align}
The Hilbert space $ \thaux=L^{2}(\R^{3},\ud\Tta\ud\Xi\ud x) $ is endowed with the positive definite inner product\reff{eq:thaux inner 2}.  Thus\reff{eq:ga case 2b} is mapped into
\begin{equation}\label{eq:tga 2b}
\int\! \ud\lm^{1}\ud\lm^{2}\,\ud\Tta\ud\Xi\ud x\:\dfrac{\sinh\big[\half\lm^{2}\kp g(x)\big]}{\half\lm^{2}\kp g(x)}\;{\wt{\psi}^{*}(\Tta,\Xi,x)}\,\wt{\chi}(\Tta+\bt,\Xi+\lm^{2},x)\ ,
\end{equation}
or written in a more familiar way
\begin{equation}\label{eq:tga 2b clear}
\int\! \ud\bt\ud\lm^{2}\,\ud\Tta\ud\Xi\ud x\:{\wt{\psi}^{*}(\Tta,\Xi,x)}\,\wt{\chi}(\Tta+\bt,\Xi+\lm^{2},x)\ ,
\end{equation}
where the definition of $ \bt $, Eq.\reff{eq:bt def}, was used. Allowing $\lm^{a} $ to take values over the whole real  line, one has that $ \bt\in(-\infty,\infty) $ and the expression\reff{eq:tga 2b clear} is equivalent to\reff{eq:tga i}. Therefore, the rescaling at the quantum level has explicitly been undone. The system was mapped to that in which the scaling functions are the identity.

RAQ in $ \thaux $ can now be carried out as at the end of the previous subsection. Choosing smooth functions of compact support on $ \R^{3}=\{(\Tta, \Xi, x)\}  $ as the test state space, on which\reff{eq:tga 2b clear}  is well defined,  one can verify that the averaging procedure projects out the gauge dependence on the wave functions leaving the physical Hilbert space $ \hfis=L^{2}(\R,\ud x) $.

\section{Final comments}
\label{sec:final comm}
In this article, we have investigated RAQ under a specific family of classically equivalent systems with two rescaled momentum-type constraints. The parametrised family of real-valued scaling functions that has been introduced contains three representative outcomes at the level of the gauge algebra structure. Depending on the values taken by the parameters, either the original abelian gauge algebra: $(i)$ is maintained, $(ii)$ is turned into a gauge algebra that presents nonunimodular behaviour with nonconstant gauge invariant structure functions, or $(iii)$ becomes a full open gauge algebra with artificially constructed structure functions --that is,  with structure functions that fully depend on the unreduced configuration space $ \R^{3} $. 

The RAQ of cases $(i)$ and $(ii)$ above was analysed in full. Each one of them was mapped into the case where the scaling functions are the identity function, recovering the physical Hilbert space of the unscaled constrained system. Due to the mathematical structure of the gauge algebra in case $(i)$, the group averaging technique was directly applied. In contrast to case $(i)$ when non-constant structure functions are  present, as in case $(ii)$ and $(iii)$, a `group averaging' formula was supplemented based on a particular BRST proposal. The BRST-canonical quantisation for cases $(ii)$ and $(iii)$ was presented in the Appendix~\ref{app:BRSTanalysis}. For case $(ii)$ the regularised BRST inner product yielded a `group averaging' formula with a ghost-free gauge invariant  symmetric measure which was fully implemented in the RAQ analysis. The RAQ  of a gauge system with structure functions, though artificially constructed and gauge invariant, was then successfully written down. This became the first example  of its kind known to the author were structure functions are successfully treated within this scheme. 

As for cases $ (i) $ and $ (ii) $, in case $ (iii) $ each classical constraint~\eqref{eq:M and N} is promoted to a self-adjoint operator; the self-adjointness Von Neumann's  criterion  applied to each constraint operator $ \wht{\phi}_{1}:=-if(x) e^{\kp_{1}\vfi}\partial_{\tta} $ and $
\wht{\phi}_{2}:= -ig(x)e^{\kp_{2}\tta}\partial_{\vfi} $ gives as a result that each pair of deficiency indices $ (n_{+},n_{-}) $ is $ (0,0) $. Therefore, the constraint operators generate an action on the auxiliary Hilbert space through unitary operators.  In this construction there are 
neither ordering issues nor anomalies at the level of the quantum gauge algebra,
\begin{equation*}
\big[\,\wht{\phi}_{1},\wht{\phi}_{2}\,\big]=i\f{1}{2}{a}(q)\wht{\phi}_{a}\ ,
\end{equation*}
this commutation relation being a consequence of the the specific form of $ M(\vfi,x) $ and $ N(\tta,x) $, see Eqs.~\eqref{eq:M and N}. Nevertheless,  the most challenging difficulties one faces in case $ (iii) $ are at the level of the BRST regularised inner product \eqref{eq:reg inner}, which proved to be useful in cases $ (i) $ and $(ii)$ where the rescaling process was undone at a quantum level. In case $ (iii) $, due to the presence of all coordinates in the structure functions $ \f{a}{b}{c}(q) $, this expression unwieldily involves fermionic momenta in its measure, \cf\reff{eq:E of t expansion}; the ghost structure does not seem to reduce to a manageable integral, hence a ghost-free measure to be used in a `group averaging' formula that can act on $ \haux $ is unclear.  Even if this involved step is overcome, we are left with the issue of arriving at the right representation of the corresponding gauge group elements on $ \haux $, such that the corresponding group averaging ansatz can be mapped, if possible, to the unscaled constraint system $ p_{\tta}\approx 0 $ and $ p_{\vfi}\approx 0 $. 

Notice that none of the three cases introduced in Sec.~\ref{sec:scaling family} necessitated an asymptotic analysis of the corresponding scaling functions, the current scaling functions do not impose any obstacle in the self-adjointness of constraints, hence the scaling functions fall into the type-I category in the classification of Paper I.

\section*{Acknowledgements}
The present work would have not been achieved without the influence of Jorma Louko, whose suggestions were crucial. The author also thanks Alicia L\'{o}pez-Osorio for her feedback after proofreading the first version of this paper. This work has been supported by CONACyT (Mexico).

\appendix

\section{Canonical BRST analysis}
\label{app:BRSTanalysis}

This Appendix is a generalisation of the Appendix A in Paper I. The canonical BRST analysis of the constraint system\reff{eq:rcs} is achieved. A `group averaging' ansatz for systems that include nonvanishing gauge invariant structure functions at the level of the gauge algebra is derived. In the quantum mechanical analysis, the issues of operator domains are left to be specified once a genuine Hilbert space is introduced in Sec.~\ref{sec:as}.

A detailed exposition of the BRST formalism is presented in Ref.\citte{henbk}, whose conventions are adopted in this Appendix.

\subsection{Classical BRST formalism}
Let $ (q^{i}) $ and $ (p_{i}) $ denote, respectively, the coordinates $ (\tta,\vfi,x) $ and momenta $ (p_{\tta},p_{\vfi}, p_{x}) $. The super phase space $ \Gm_{\!\!\trm{ext}} $,  is obtained by trivially enlarging the phase space $ \Gm=\{(p,q)\}$ with the canonical pairs $(\lm^{a},\pi_{a})$, $(C^{a},\bar{\ro}_{a})$, $(\bar{C}_{a},\ro^{a})$. Besides the constraints\reff{eq:rcs}, new constraints are defined on $ \Gm_{\!\!\trm{ext}} $, conjugate momenta to the Lagrange multipliers $ \lm^{a} $ are set to vanish: $ \pi_{a}\approx 0 $.   Together $ \phi_{a} $ and $ \pi_{a} $ form a first-class set of bosonic constraints. The non-trivial part of the super-symplectic structure on $ \Gm_{\!\!\trm{ext}} $ reads
\begin{equation}\label{eq:ext 2form}
\begin{aligned}
\{q^{i}, p_{j}\} & =\dlt^{i}_{j}, \quad\; \{\lm^{a}, \pi_{b}\} =\dlt^{a}_{b}\ \ \ \, \trm{(Bosons)}\\
\{C^{a}, \bar{\ro}_{b}\} & = -\dlt^{a}_{b},\ \{\bar{C}_{a},\ro^{b}\}=-\dlt^{a}_{b}\ \hfill\trm{(Fermions)}.
\end{aligned}
\end{equation}

The pairs of fermionic variables $ (C^{a},\bar{\ro}_{a}) $ and $ (\bar{C}_{a},\ro^{a}) $ belong, respectively, to the non-minimal and minimal sectors of the theory. In the conventions of Refs.\citte{henbk, henrep}, all fermionic momenta $ (\bar{\ro},\ro) $ are imaginary and their conjugate pairs $  (C,\bar{C})$, as well as all bosonic variables, are real\citte{Note1}
 . Ghost number zero is attached to all bosonic variables, ghost number $ +1 $ to the variables $ C $ and $ \ro $, and ghost number $ -1 $ is attached to $\bar{C}$ and $ \bar{\ro}$. Ghost numbering is generalised to any monomial on $ \Gm_{\!\!\trm{ext}} $ by the arithmetic addition of the ghost number of each factor; it is only the sum of monomials with the same ghost number which specifies a polynomial of definite ghost number.

 The BRST generator associated to the system of constraints\reff{eq:rcs} is
\begin{equation}\label{eq:omnm+omm}
\Om=\Om^{\!\trm{min}}-i\ro^{a}\pi_{a}
\end{equation}
where the minimal sector of the BRST generator is
\begin{equation}\label{eq:omm}
\Om^{\!\trm{min}}=M(q)\,p_{\tta}\,C^{1}+N(q)\,p_{\vfi}\,C^{2}+\half\f{a}{b}{c}(q)\,C^{a}C^{b}\bar{\ro}_{c}\ .
\end{equation}

The BRST generator $ \Om $ is real, of ghost number $ +1 $ and nilpotent in the PB sense: $ \{\Om,\Om\}=0 $. Notice that $ \Om $ contains structure functions up to rank 1. Since the  Latin indices take only two different values, 1 and 2, one has that  the second (or higher) order structure functions automatically vanish. It is only the number of constraints $ \phi_{a} $ which determines the rank of the set of constraints in the BRST sense. The \emph{a posteriori} introduced constraints, $ \pi_{a}\approx 0$, are abelian and have vanishing PB with the minimal sector of constraints.

If more than two momentum constraints were present, the possibility of having nonvanishing higher order structure functions in the BRST generator becomes authentic; however, provided that  the constraints are all linear in momenta, the following theorem rules out such possibility.

\begin{theorem}\label{teo:om for linear momentum constraints}
Let $(q^{i},p_{i})$ be the coordinates of a representative point on a phase space on which a set of regular and irreducible constraints, linear in momenta, is defined. Then, the classical minimal BRST charge $\Om^{\!\trm{\emph{min}}}$ can be taken to be linear in the momenta $(p_{i},\bar{\ro}_{a})$.
\end{theorem}
A generalisation of this theorem, to cases where reducible linear-momentum constraints take place, can be found as Proposition 1 in Ref.\citte{fer93}, its proof can easily be specialised to prove the validity  of the Theorem~\ref{teo:om for linear momentum constraints}. It is remarkable that even when the gauge algebra of constraints may contain structure functions of configuration variables, the cumbersome complete BRST generator\citte{henrep} can be truncated to include the zero and first order structure functions only without losing any gauge information.

\subsection{BRST quantisation}
 
As in Paper I, a representation in which the wave functions depend on the bosonic coordinates $(q,\lm)$ and the fermionic momenta $(\bar{\ro},\ro)$ is chosen. A wave function can be expanded in the fermionic variables as 
\begin{align}\label{eq:brst wave fncs}
\Psi(q, \lm,\bar{\ro},\ro) = &\ \psi(q,\lm)+\Psi^{a}(q,\lm)\,\bar{\ro}_{a}+\half\Psi^{[ab]}(q,\lm)\,\bar{\ro}_{a}\bar{\ro}_{b}+\half\Psi^{[ab]}_{c}\,\bar{\ro}_{a}\bar{\ro}_{b}\ro^{c}\nonumber\\
&+\half\Psi^{a}_{[bc]}(q,\lm)\,\bar{\ro}_{a}{\ro}^{b}\ro^{c}+\Psi_{a}(q,\lm)\,\ro^{a} +\half\Psi_{[ab]}(q,\lm)\,{\ro}^{a}\ro^{b}\nonumber\\
&+\Psi^{12}_{12}(q,\lm)\,\bar{\ro}_{1}\bar{\ro}_{2}\,{\ro}^{1}\ro^{2}\ .
\end{align}
with the coefficients being complex-valued functions. The action of the fundamental operators on wave functions\reff{eq:brst wave fncs} read
\begin{subequations}\label{eq:basic brst operators action}
\begin{align}
\wht{q}^{\; i}\,\Psi & := q^{i}\Psi\ , & \wht{p}_{i}\,\Psi &:=-i\dfrac{\partial\Psi}{\partial q^{i}}\ ,\label{eq:system3; q,p action} \\
\wht{\lm}^{a}\,\Psi & := \lm^{a}\Psi \ , & \wht{\pi}_{a}\,\Psi &:=-i\frac{\partial\Psi}{\partial\lm^{a}}\ , \label{eq:system3; lm,pi action}\\
\wht{C}^{a}\,\Psi & := -i\dlpartial{\Psi}{\bar{\ro}_{a}}\ , &  \wht{\!\bar{\ro}}_{a}\,\Psi &:=\bar{\ro}_{a}\Psi \ ,\label{eq:system3; c,bro action}\\
\wht{\!\bar{C}}_{a}\,\Psi & :=-i\dlpartial{\Psi}{\ro^{a}}\ , & \wht{\ro}^{\,a}\,\Psi & := \ro^{a}\,\Psi\ , \label{eq:system3; bc,ro action}
\end{align}
\end{subequations}
where $ \wht{p}_{i} $ denotes the momentum operators $ \wht{p}_{\tta} $, $ \wht{p}_{\vfi} $ and $ \wht{p}_{x} $. The superscript $ \ell $ on the fermionic derivative stands for left derivative. This choice is compatible with the basic graded commutation relations (\cf(\ref{eq:ext 2form})):
\begin{subequations}\label{eq:graded comm rels}
\begin{align}
[\wht{q}^{\;i},\wht{p}_{j}] & =i\dlt^{i}_{j}\ , \,\qquad[\wht{\lm}^{a},\wht{\pi}_{b}\,]=i\dlt^{a}_{b}\ ,\label{eq:system3; graded comm rels bosons}\\
[\wht{C}^{a},\,\wht{\!\bar{\ro}}_{b}\;]& = -i\dlt^{a}_{b}\ ,\quad  [\,\wht{\!\bar{C}}_{b},\wht{{\ro}}^{\,a}\,]=-i\dlt^{a}_{b}\ .\label{eq:system3; graded comm rels fermions}
\end{align}
\end{subequations}

In the canonical BRST formalism the physical quantum states satisfy 
\begin{subequations}
\label{eq:brsc-and-ghnopc}
\begin{align}
\label{eq:brsc}
\wht{\Om} \Psi &=0, 
\\
\label{eq:ghnopc}
\wht{N}_{\mc{G}}\Psi &=0,  
\end{align}
\end{subequations}
where the ghost number operator $\wht{N}_{\mc{G}}$ is defined by
\begin{equation}\label{eq:ggnop}
\wht{N}_{\mc{G}} :=i(\wht{\ro}^{\,a}\,\wht{\!\bar{C}}_{a}-\,\wht{\!\bar{\ro}}^{\,a}\wht{C}_{a})\ .
\end{equation}
The construction of the BRST quantum operator $ \wht{\Om} $ corresponding to\reff{eq:omnm+omm} involves two sources of ordering issues: (1) in the definition of $ \wht{\phi}_{a} $, and (2) within the terms which contain structure functions. The first issue is solved by choosing the following symmetric order:
\begin{subequations}\label{eq:cons ops}
\begin{align}
\wht{\phi}_{1} & := -i\big(M(q)\partial_{\tta}+\half(\partial_{\tta}M)\big)\ , \label{eq:fi1 operator}\\
\wht{\phi}_{2} & := -i\big(N(q)\partial_{\vfi}+\half(\partial_{\vfi}N)\big) \ .   \label{eq:fi2 operator}
\end{align}
\end{subequations}
The second issue finds a resolution when all fermionic momenta $ (\bar{\ro},\ro) $ appear at the left of the fermionic coordinates $ (C,\bar{C}) $, that is, 
\begin{equation}\label{eq:om}
\wht{\Om}\equiv\wht{\Om}_{L} :=\,  \big(\wht{\phi}_{a}-\dfrac{i}{2}\f{a}{b}{b}(q)\big)\,\wht{C}^{a}+\half\f{a}{b}{c}(q)\,\,\wht{\!\bar{\ro}}_{c}\,\wht{C}^{a}\,\wht{C}^{b}-i\wht{\ro}^{\,a}\,\wht{\pi}_{a}.
\end{equation}
An equivalent resolution is when all fermionic momenta $ (\bar{\ro},\ro) $ appear at the right of the fermionic coordinates $ (C,\bar{C}) $, that is,
\begin{equation*}
\wht{\Om}_{R} : =\,  \wht{C}^{a}\,\big(\wht{\phi}_{a}+\dfrac{i}{2}\f{a}{b}{b}(q)\big)+\half\f{a}{b}{c}(q)\,\wht{C}^{a}\,\wht{C}^{b}\,\,\wht{\!\bar{\ro}}_{c}-i\wht{\pi}_{a}\wht{\ro}^{\,a}\ .
\end{equation*}
With this ordering of factors, and $ \f{a}{b}{c} $ structure constants, the physical state conditions Eqs.\reff{eq:brsc-and-ghnopc} on zero ghost numbered states\reff{eq:triv-phy-state}  (see below) are translated into the Dirac physical state condition for non-unimodular groups reported in Refs.\citte{giu99a, Louko:2004zq}. When $ \f{a}{b}{b}\neq 0 $ the Dirac physical state condition is generalised\citte{mar93a,mar94,mar94a,mar95,bat95}.

Let $ X $ be any state, the so-called BRST gauge transformation $ \Psi\mapsto\Psi':=\Psi+\wht{\Om}X $ preserves the condition\reff{eq:brsc} since $ \wht{\Om}^{2}=0 $. Moreover, if $ X $ has ghost number $ -1 $, a gauge transformation also preserves the condition\reff{eq:ghnopc}. A gauge transformation in which $ X $ has ghost number $-1 $ hence takes physical states to physical states.

As usual, two BRST general  states\reff{eq:brst wave fncs} are paired using the nondefinite sesquilinear form
\begin{equation}\label{eq:brs sesqui form}
\brs{\Psi}{\Ups}:=c\int \ud^{2}\lm\,\ud^{3} q\,\ud^{2}\bar{\ro}\,\ud^{2}\ro\;\Psi^{*}(q,\lm,\bar{\ro},\ro)\Ups(q,\lm,\bar{\ro},\ro)\ , \ c\in\C.
\end{equation}
For any value of the complex constant  $ c $, the sesquilinear form\reff{eq:brs sesqui form}  has the following properties. First, it is compatible with the reality conditions of the classical variables --that is, fermionic ghost momentum operators are anti-hermitian and the rest of fundamental operators in\reff{eq:basic brst operators action} are hermitian. Second, $ \wht{\Om} $ is hermitian in the sense of\reff{eq:brs sesqui form}. Third, the sesquilinear form $ \brs{\cdot}{\cdot} $ on physical states depends on the states only through their gauge-equivalence class. Fourth, letting $ c $ be real, the indefinite sesquilinear form  $ \brs{\cdot}{\cdot} $ becomes hermitian; in particular $ c=-1 $ is chosen to link the regularised inner product (see below\reff{eq:reg inner}) and a group averaging formula.

\subsection{Averaging ansatz}

There is a special type of zero ghost numbered states that trivially satisfy the physical state conditions\reff{eq:brsc-and-ghnopc}, these are
\begin{equation}\label{eq:triv-phy-state}
\Psi = \psi(q).
\end{equation}
The non-minimal sector of the BRST operator\reff{eq:om} annihilates these states due to their $ \lm $-independence. From now on, the attention is focused on those physical states with neither fermions nor Lagrange multipliers\reff{eq:triv-phy-state}.

The sesquilinear form\reff{eq:brs sesqui form} is not well defined when evaluated on states\reff{eq:triv-phy-state}. As in Paper I, the  following regularised sesquilinear form is considered:
\begin{align}\label{eq:reg inner}
\brs{\psi}{\chi}^{\vro} & := \brs{\psi}{\wht{V}\chi} \nonumber\\
&=\ -\int \ud^{2}\lm\,\ud^{3} q\,\ud^{2}\bar{\ro}\,\ud^{2} \ro\;\psi^{*}(q)\big[\exp{\big(i[\,\wht{\Om},\wht{\vro}\,]\big)}\chi\,\big](q)\ , 
\end{align}
where $ \wht{V}:=\exp(i[\wht{\Om},\wht{\vro}\,]) $ may be turned into a unitary operator with respect to $ \brs{\cdot}{\cdot} $ once all the fermionic momenta are integrated out. 

Consider the anti-hermitian gauge fixing fermion $ i\wht{\vro}:=-\wht{\lm}^{a}\,\,\wht{\bar{\!\ro}}_{a} $. By direct calculation, it can be proved that on general BRST states\reff{eq:brst wave fncs}, one has
\begin{equation}\label{eq:om,K}
i[\,\wht{\Om},\wht{\vro}\,]\Psi=(i\lm^{a}\,\wht{\phi}'_{a}-iu_{a}\wht{C}^{a}+\ro^{a}\bar{\!\ro}_{a})\Psi
\end{equation}
where $ u_{a}:=\f{b}{a}{c}\,\lm^{b}\,\bar{\!\ro}_{c} $ and $ \wht{\phi}'_{a} := \wht{\phi}_{a}-\half i\f{a}{b}{b}(q)$. In order to evaluate $ \exp(i[\wht{\Om},\wht{\vro}\,]) $ on physical states\reff{eq:triv-phy-state}, the following definitions are found useful:
\begin{equation}\label{eq:A-B operators}
\wht{A}:=\ro^{a}\,\,\bar{\!\ro}_{a}\,\I,\quad \wht{B}:=-iu_{a}\,\wht{C}^{a}\ .
\end{equation}
Hence $ i[ \wht{\Om}, \wht{\vro}\,]=i\lm^{a}\,\wht{\phi}{}'_{a}+\wht{A}+\wht{B} $. On wave functions\reff{eq:brst wave fncs} the following commutators are satisfied:
\begin{align}
[\,i\lm^{a}\,\wht{\phi}{}'_{a}, \wht{A}+\wht{B}\,] & =\lm^{c}(\partial_{c}\,u^{a}_{\ph{a}b})\,\,\bar{\!\ro}_{a}\,\wht{C}^{b}  \label{eq:lmfi,A+B}\\
[\,\wht{A},\,\wht{B}\,] & =-\,\,\bar{\!\ro}_{a} \,u^{a}_{\ph{a}b}\,\ro^{b}\label{eq:A,B}
\end{align}
with $ u^{a}_{\ph{a}b}=u^{a}_{\ph{a}b}(q,\lm):=\f{c}{b}{a}(q)\lm^{c}$ and the notation $ \partial_{a} $ in this Appendix refers to $-iM(q)\partial_{\tta} $ and $ -iN(q)\partial_{\vfi} $ when $ a=1 $ and $ a=2 $, respectively. 
Define the following operator:
\begin{equation}\label{eq:E(t)}
\wht{E}(t):=\exp\big(it[\,\wht{\Om},\wht{\vro}\,]\big)\exp\big(\!-t\wht{B}\big)\ .
\end{equation}
Due to the property $ \wht{C}^{a}\psi=0 $ on physical states\reff{eq:triv-phy-state}, one has the following chain of equalities $\wht{E}(1)\psi=\exp\big(i[\,\wht{\Om},\wht{\vro}\,]\big)\,\exp\big(\!-\wht{B}\big)\psi=\exp\big(i[\,\wht{\Om},\wht{\vro}\,]\big)\,\psi$. Differentiating\reff{eq:E(t)} with respect to $ t $, one obtains
\begin{equation}\label{eq:E diffeq}
\dfrac{\ud\wht{E}}{\ud t}=\wht{E}(t)\,\wht{r}(t)\ ,
\end{equation}
with $  \wht{r}(t)\equiv \exp({t\wht{B}})\,\big(i\lm^{a}\wht{\phi}{}'_{a}+\wht{A}\,\big)\exp(\!-t\wht{B}) $. Equation\reff{eq:E diffeq} may be solved by iteration of the corresponding integral equation
\begin{equation}\label{eq:E of t first integral}
\wht{E}(t)=\I+\int_{0}^{t}\ud t_{1}\;\wht{E}(t_{1})\,\wht{r}(t_{1})\ ,
\end{equation}
where the boundary condition $\displaystyle{\lim_{t\to 0}\wht{E}(t)=\I}  $ has been introduced, it results in
\begin{align}\label{eq:E of t expansion}
\wht{E}(t) = & \ \I+\int_{0}^{t}\ud t_{1}\;\wht{r}(t_{1})+\int_{0}^{t}\ud t_{1}\int_{0}^{t_{1}}\ud t_{2}\;\wht{r}(t_{2})\wht{r}(t_{1})+\ldots\notag \\
 & \ + \int_{0}^{t}\ud t_{1}\int_{0}^{t_{1}}\ud t_{2}\,\cdots\int_{0}^{t_{n-1}}\ud t_{n}\;\wht{r}(t_{n})\wht{r}(t_{n-1})\cdots\wht{r}(t_{1})+ \ldots\ .
\end{align}
Here the ordering of the operators is important and a `$ t $-ordering' symbol may be used. Instead,  using the Baker-Hausdorff lemma, a direct calculation shows that
\begin{equation}\label{eq:r hat}
\wht{r}(t)=i\lm^{a}\wht{\phi}{}'_{a}-\bar{\!\ro}_{a}\big(e^{-t\mb{u}}\big)^{a}_{\ph{a}b}\,\ro^{b}-\lm^{c}\,\bar{\!\ro}_{a}(\Dlt_{c}(t))^{a}_{\ph{a}b} \,\wht{C}^{\,b} \ ,
\end{equation}
with
\begin{equation}\label{eq:dlt c}
(\Dlt_{c}(t))^{a}_{\ph{a}b}  \equiv\partial_{c}(t\mb{u})^{a}_{\ph{a}b}+\dfrac{1}{2!}[\partial_{c}(t\mb{u}),(t\mb{u})]^{a}_{\ph{a}b}+\dfrac{1}{3!}\big[[\partial_{c}(t\mb{u}),(t\mb{u})],(t\mb{u})\big]^{a}_{\ph{a}b}+\cdots\ ,
\end{equation}
where $ \mb{u} $ is the matrix whose elements are $ u^{a}_{\ph{a}b}(q,\lm) $ and the square brackets refer to the usual commutator of matrices $ [\mb{u},\mb{v}]^{a}_{\ph{a}b}:= u^{a}_{\ph{a}c}\,v^{c}_{\ph{a}b}-v^{a}_{\ph{a}c}u^{c}_{\ph{c}b}$. From the expression\reff{eq:r hat}, one can see that for\reff{eq:E of t expansion} the order of  operators becomes irrelevant if the structure functions depend at most on the true degree of freedom, that is, $ \f{a}{b}{c}=\f{a}{b}{c}(x) $. Indeed, in such cases $(\Dlt_{c}(t))^{a}_{\ph{a}b} \equiv0$ and it follows $ [\,\wht{r}(t),\wht{r}(t')\,]=0 $. Each integrand on the RHS of\reff{eq:E of t expansion} becomes symmetric in the parameters $ t_{i} $, and one can convince oneself that the $ n$th  integral satisfies the following identity:
\begin{equation*}
n! \int_{0}^{t}\ud t_{1}\int_{0}^{t_{1}}\ud t_{2}\,\cdots\int_{0}^{t_{n-1}}\ud t_{n}\;\wht{r}(t_{n})\wht{r}(t_{n-1})\cdots\wht{r}(t_{1})=\left[\int_{0}^{t}\ud t'\;\wht{r}(t')\right]^{n}\ .
\end{equation*}
Hence
\begin{equation}\label{eq:F(t) solution}
\wht{E}(t)=\,\exp\left({\int_{0}^{t}\ud t'\,\wht{r}(t')}\right)\ \quad \big(\f{a}{b}{c}=\f{a}{b}{c}(x)\big) \ ,
\end{equation}
making use of\reff{eq:r hat}, a direct integration gives
\begin{equation}\label{eq:F(1) explicit}
\wht{E}(1)\psi  =\  \exp\left[i\lm^{a}\wht{\phi}{}'_{a}-
 \,\,\bar{\!\ro}_{a}\left(\dfrac{\I-e^{-\mb{u}}}{\mb{u}}\right)^{a}_{\ph{a}b}\ro^{b}\right]\psi\ .
\end{equation}

Therefore, under the assumption that $ f_{ab}^{\ph{ab}c} $ only  depends on $ x $, one has that on physical states\reff{eq:triv-phy-state}
\begin{equation}\label{eq:exp i om,K}
\wht{V}\psi=\exp\big(i[\,\wht{\Om},\wht{\vro}\,]\big)\psi=\exp\left( \,\bar{\!\ro}_{a}\Big(\dfrac{e^{-\mb{u}}-\I}{\mb{u}}\Big)^{a}_{\ph{a}b}\,\ro^{b}+i\lm^{a}\wht{\phi}{}'_{a}\right)\psi\ ,
\end{equation}
where the $ 2\times 2$ matrix $ \mb{u} $ explicitly reads
\begin{equation}\label{eq:u matrix case i}
\mb{u}(x,\lm)=
\begin{pmatrix} 
f_{21}^{\ph{21}1}(x)\lm^{2} & f_{12}^{\ph{21}1}(x)\lm^{1}\\ 
f_{21}^{\ph{21}2}(x)\lm^{2} & f_{12}^{\ph{21}2}(x)\lm^{1}
\end{pmatrix}\ 
\end{equation}
and
\begin{equation}\label{eq:exp(1-e(-u))}
\left(\dfrac{\I-e^{-\mb{u}}}{\mb{u}}\right)^{a}_{\ph{a}b} \equiv \dlt^{a}_{\ph{a}b}-\dfrac{1}{2!}u^{a}_{\ph{a}b}+\dfrac{1}{3!}u^{a}_{\ph{a}c}u^{c}_{\ph{a}b}-\dfrac{1}{4!}u^{a}_{\ph{a}c}u^{c}_{\ph{a}d}u^{d}_{\ph{a}b}+\cdots \ .
\end{equation}

Inserting the expression\reff{eq:exp i om,K} into the regularised BRST inner product\reff{eq:reg inner} and solving the Gaussian integral in the Grassmann variables,
\begin{equation}\label{eq:final reg inner}
\reg{\psi}{\chi} = \int \ud^{2}\lm\,\ud^{3} q\,\vert j_{0}(\mb{u}(x,\lm))\vert\;\psi^{*}(q)\big[\exp{\big(i\lm^{a}\wht{\phi}_{a}\big)}\chi\,\big](q)\ ,
\end{equation}
where the gauge invariant and positive definite factor $ \vert j_{0}(\mb{u}(x,\lm))\vert $ is defined as
\begin{equation}\label{eq:symm measure}
\vert j_{0}(\mb{u}(x,\lm))\vert:=\det\left[\dfrac{e^{\mb{u}(x,\lm)/2}-e^{-\mb{u}(x,\lm)/2}}{\mb{u}(x,\lm)}\right]\ .
\end{equation}

The matrix $ \mb{u}(x,\lm) $ defined in\reff{eq:u matrix case i} vanishes if the the gauge group shows unimodular behaviour, $ f_{ab}^{\ph{ab}b}(x)=0 $. In this limit, the positive quantity $ \vert j_{0}(\mb{u})\vert $ becomes the identity. Therefore, if $ f_{ab}^{\ph{ab}b}(x)=0 $, the structure of the regularised BRST inner product\reff{eq:final reg inner} coincides with the group averaging formula of a unimodular group\citte{ash95} provided that the exponential of the constraint operators can be properly interpreted as unitary operator on some auxiliary Hilbert space and that, simultaneously, the integrals over the Lagrange multipliers converge in some sense. If solely structure constants are present in the gauge algebra and there is non-unimodularity of the gauge group, the measure $ \vert j_{0}(\mb{u})\vert \ud^{2}\lm $ coincides with the symmetric measure in the adjoint representation of the corresponding gauge group\citte{giu99a, rosbk} (see also Appendix \ref{app:gauge group}). Eq.\reff{eq:symm measure} duly reduces to the group averaging ansatz given in Ref.\citte{giu99a}
 
\section{The gauge group at \texorpdfstring{$ x $}{}}\label{app:gauge group}

In this Appendix, the basic properties of the gauge group generated by\reff{eq:M and N} in the case $ \kp_{2}=0 $ and $ \kp_{1}\equiv\kp\neq 0 $ are placed.

At each point $ x $, the gauge group is the subgroup $ B(2,\R) $ of $ GL(2,\R) $, that is, upper triangular with matrices such that $ g_{11}>0 $ and $ g_{22}=1 $. This is a two-dimensional, non-abelian group with $ x $- dependent nonunimodular behaviour, that is, $ \f{a}{b}{b}(x) \neq 0$.

The Lie algebra $ \mf{b}(2,\R) $, at each point $ x $,  is spanned by the following matrices with $ x $ dependent coefficients:
\begin{equation}\label{eq:gl2r}
T_{1}(x) := \kp g(x)\begin{pmatrix} 0 & 1\\  0&0  \end{pmatrix}\ , \quad T_{2}(x) := \kp g(x)\begin{pmatrix} 1 & 0\\  0&0  \end{pmatrix} \ , 
\end{equation}
which obey the algebra
\begin{equation}\label{eq:antialgebra}
\big[ T_{1}(x),T_{2}(x)\big]=-\kp g(x)\, T_{1}(x)\ .
\end{equation}
The mapping $ \phi_{a}\mapsto T_{a} $ becomes an anti-homomorphism of Lie algebras at each point $ x $, \cf\reff{eq:comm of constraints ii}.
Elements of $ B(2,\R) $ can be written as the exponential of $ \lm^{a}T_{a}(x) $, with $ \lm^{a}\in\R $, 
\begin{equation}\label{eq:exp1}
\exp\left(\lm^{a}T_{a}(x)\right)=\begin{pmatrix} e^{\lm^{2}\kp g(x)} & \dfrac{\lm^{1}}{\lm^{2}}(e^{\lm^{2}\kp g(x)}-1)\\  0&1  \end{pmatrix}=:g(\lm^{a})\ \in\ B(2,\R)\ ,
\end{equation}
from which $ g^{-1}(\lm^{a})=g(-\lm^{a}) $ and $ \displaystyle{\lim_{\lm^{a}\to 0} g( \lm^{a})=\I}$. 

An equivalent way to write elements of $ B(2,\R) $ is through the factorisation
\begin{equation}\label{eq:exp3}
g=\exp\left(\dfrac{\lm^{2}}{2}T_{2}(x)\right)\exp\big(\bt T_{1}(x)\big)\exp\left(\dfrac{\lm^{2}}{2}T_{2}(x)\right)\ ,
\end{equation}
with $ \bt $ uniquely defined by
\begin{equation}\label{eq:bt def}
\bt:=\lm^{1}\, \dfrac{\sinh\left(\half\kp g(x)\lm^{2}\right)}{\half\kp g(x)\lm^{2}}\ .
\end{equation}

A direct calculation shows that $ g^{-1}\ud g= W^{a}(x)T_{a}(x) $, where the left-invariant 1-forms $ W^{a}(x) $ are given by
\begin{subequations}\label{eq:lif}
\begin{align}
W^{1}(x) &:= \ \dfrac{1-e^{\lm^{2}\kp g(x)}}{\lm^{2}\kp g(x)}\,\ud\lm^{1}+\dfrac{w(\lm^{a},x)}{\kp g(x)}\, e^{-\lm^{2}\kp g(x)}\,\ud\lm^{2} \ , \label{eq:W1}\\
W^{2}(x) &:= \ \ud\lm^{2} \label{eq:W2}\ ,
\end{align}
\end{subequations}
with $ w(\lm^{a},x) $ a function whose specific dependence is irrelevant in the present analysis, henceforth it is denoted as $ w $; a similar analysis shows  that $\ud g g^{-1}= \dot{W}^{a}(x)T_{a}(x) $, from which the following right-invariant 1-forms can be read:
\begin{subequations}\label{eq:rif}
\begin{align}
\dot{W}^{1}(x) &:= \ \dfrac{e^{\lm^{2}\kp g(x)}-1}{\lm^{2}\kp g(x)}\,\ud\lm^{1}+\left[\dfrac{\lm^{1}}{\lm^{2}}\left(1-e^{\lm^{2}\kp g(x)}\right)+\dfrac{w}{\kp g(x)}\right]\,\ud\lm^{2} \ , \label{eq:W1dot}\\
\dot{W}^{2}(x) &:= \ \ud\lm^{2} \label{eq:W2dot}\ .
\end{align}
\end{subequations}
From the expressions\reff{eq:lif} and\reff{eq:rif} the left- and right-invariant measures forms can be constructed,
\begin{subequations}\label{eq:lrim}
\begin{align}
(\ud_{L}g) (x):= &\  W^{1}(x)\x W^{2}(x)=\dfrac{1-e^{-\lm^{2}\kp g(x)}}{\lm^{2}\kp g(x)}\,\ud\lm^{1}\ud\lm^{2} \label{eq:left m}\ ,\\
(\ud_{R}g)(x):= & \ \dot{W}^{1}(x)\x \dot{W}^{2}(x)=\dfrac{e^{\lm^{2}\kp g(x)}-1}{\lm^{2}\kp g(x)}\,\ud\lm^{1}\ud\lm^{2} \label{eq:right m}\ .
\end{align}
\end{subequations}

The adjoint action of the group $ B(2,\R) $ on its Lie algebra $ \mf{b}(2,\R) $ reads $ (\Ad{g}\,T_{1})(x) = (gT_{1}g^{-1})(x)$ $=e^{\lm^{2}\kp g(x)}\,T_{1}(x)$ and $ (\Ad{g}\,T_{2})(x)=(gT_{2}g^{-1})(x)=\dfrac{\lm^{1}}{\lm^{2}}(1-e^{\lm^{2}\kp g(x)})\,T_{1}(x)+T_{2}(x)$. The modular function at each point $ x $ can then be obtained $ \Dlt(g)=\det(\Ad{g})=e^{\lm^{2}\kp g(x)} $. The gauge invariant symmetric measure, invariant under $ g\mapsto g^{-1} $, is
\begin{equation}\label{eq:symm measure b2r}
(\ud_{0}g)(x)=([\Dlt(g)]^{1/2}\,\ud_{L}g)(x)=([\Dlt(g)]^{-1/2}\,\ud_{R}g)(x)=\dfrac{\sinh[\half\lm^{2}\kp g(x)]}{\half\lm^{2}\kp g(x)}\ud\lm^{1}\ud\lm^{2}\ .
\end{equation}

\section*{References}
{\singlespace
}

\end{document}